\documentclass[aps,prd,twocolumn,superscriptaddress]{revtex4-1}
\usepackage{graphicx}
\usepackage{bm}
\usepackage{amsmath}
\usepackage{amssymb}
\usepackage{amsthm}

\newcommand{\be}{\begin{equation}}
\newcommand{\ee}{\end{equation}}
\newcommand{\bea}{\begin{eqnarray}}
\newcommand{\eea}{\end{eqnarray}}
\newcommand{\bdm}{\begin{displaymath}}
\newcommand{\edm}{\end{displaymath}}
\newcommand{\beas}{\begin{eqnarray*}}
\newcommand{\eeas}{\end{eqnarray*}}

\begin{document}
\title{Fundamental Cosmology from Precision Spectroscopy: I. Varying Couplings}
\author{A. C. O. Leite}  
\email[]{up090308020@alunos.fc.up.pt}
\affiliation{Faculdade de Ci\^encias, Universidade do Porto, Rua do Campo Alegre, 4150-007 Porto, Portugal}
\affiliation{Centro de Astrof\'{\i}sica, Universidade do Porto, Rua das Estrelas, 4150-762 Porto, Portugal}
\author{C. J. A. P. Martins}
\email[]{Carlos.Martins@astro.up.pt}
\affiliation{Centro de Astrof\'{\i}sica, Universidade do Porto, Rua das Estrelas, 4150-762 Porto, Portugal}
\author{P. O. J. Pedrosa}
\email[]{ppedrosa@alunos.fc.up.pt}
\affiliation{Faculdade de Ci\^encias, Universidade do Porto, Rua do Campo Alegre, 4150-007 Porto, Portugal}
\affiliation{Centro de Astrof\'{\i}sica, Universidade do Porto, Rua das Estrelas, 4150-762 Porto, Portugal}
\author{N. J. Nunes}
\email[]{njnunes@fc.ul.pt}
\affiliation{Faculty of Sciences and Centre for Astronomy and Astrophysics, University of Lisbon, 1749-016 Lisbon, Portugal}
\date{31 August 2014}

\begin{abstract}
The observational evidence for the acceleration of the universe demonstrates that canonical theories of cosmology and particle physics are incomplete, if not incorrect, and that new physics is out there, waiting to be discovered. Forthcoming high-resolution ultra-stable spectrographs will play a crucial role in this quest for new physics, by enabling a new generation of precision consistency tests. Here we focus on astrophysical tests of the stability of nature's fundamental couplings, and by using Principal Component Analysis techniques further calibrated by existing VLT data we discuss how the improvements that can be expected with ESPRESSO and ELT-HIRES will impact on fundamental cosmology. In particular we show that a 20 to 30 night program on ELT-HIRES will allow it to play a leading role in fundamental cosmology.
\end{abstract}

\maketitle

\section{\label{intro}Introduction}

Cosmology is now a data-driven science. This is manifest in the so-called concordance model---a remarkably simple model in the sense of being able to fit observations with a small number of free parameters, though at a cost of assuming that $96\%$ of the contents of the universe are in a still unknown form (never directly observed thus far). It is thought that this dark sector has two components: a clustered one (dark matter), and a dominant unclustered one (dark energy) which is presumably responsible for the observed acceleration of the universe. Characterizing the properties of these dark components, and ultimately understanding their nature, is the key driver for modern cosmological research.

While $\Lambda$CDM provides the simplest viable possibility, it is arguably vulnerable to standard fine-tuning arguments: one needs to explain why the vacuum energy density is many orders of magnitude smaller than one would expect from particle physics based arguments. One may therefore argue that alternatives involving scalar fields, an example of which is the recently discovered Higgs field \cite{ATLAS,CMS}, may be more likely. Observationally, the main difference between the two paradigms is that in the first case the density of dark energy is always constant (it does not get diluted by the expansion of the universe) while in the second one it does change. One way to distinguish the two possibilities is to find ways to measure the dark energy density (or its equation of state) at several epochs in the universe.

Astrophysical measurements of nature's fundamental couplings  \cite{Martins:2002fm,Uzan:2010pm} can be used to constrain the properties of dynamical scalar fields that might also be responsible for the dark energy. These measurements can either be used by themselves or in combination with other cosmological datasets (such as Type Ia supernovas and the cosmic microwave background). The concept behind this method is described in \cite{Nunes:2003ff,Avelino:2006gc} (see also \cite{Parkinson:2003kf} for a more phenomenological approach). It complements other methods due to its large redshift lever arm and the fact that these measurements can be done from ground-based facilities, both in the UV/optical and the radio/mm bands.

In \cite{Amendola} we extended Principal Component Analysis (PCA, see e.g. \cite{Huterer:2002hy}) methods previously available in the published literature (for type Ia supernovae, lensing and several other contexts in cosmology) and studied the feasibility of applying them to astrophysical measurements of varying couplings---whether they are detections of variations or null results---by forecasting the number of modes of the dark energy equation of state parameter that can be well constrained by future facilities, using a combination of supernovae data and measurements of varying fundamental couplings at high redshift.

Some recent observational data suggests that the fine-structure constant $\alpha$ (a dimensionless measure of the strength of electromagnetism) was different at redshifts $z\sim2-3$, the relative variation being at the level of a few parts per million \cite{Dipole}. Various efforts to confirm or refute this result are ongoing \cite{LP1,LP2}, but a detailed answer to this important question may have to wait for the next generation of higher-resolution, more stable spectroscopic facilities. Nevertheless, and despite the fact that tests of the stability of fundamental couplings are a key science driver for future instruments, it is clear that observation time on these top facilities will be scarce, and therefore optimized observational strategies are essential.

Here we take some steps towards fully quantifying the potentialities of this method. We use currently available varying $\alpha$ measurements from VLT/UVES as a benchmark that can be extrapolated into future (simulated) datasets whose impact for dark energy characterization can be studied. We will be interested in ESPRESSO (for the VLT) \cite{ESP0}, and especially in the E-ELT's high-resolution spectrograph (ELT-HIRES) \cite{EELT,EELT2}, but our methodology is generic. In the present paper we concentrate on the varying coupling measurements themselves, while in a companion paper we will discuss in more detail the synergies between these measurements and other datasets (such as Type Ia supernovas).

In the next section we review the relevant PCA methodology and summarize and extend the results of our PCA analysis in \cite{Amendola}. Then in Sect. III we study the relevant features of the main existing database of VLT measurements of $\alpha$, allowing us to relate our theoretical PCA analysis to observationally relevant properties. Finally in Sect. IV we combine the two analysis and discuss future prospects for ESPRESSO and ELT-HIRES, and summarize our results in Sect. V.

\section{\label{methods}Theory and tools}

We will base our theoretical analysis on PCA techniques. Our formalism is described in \cite{Amendola}, to which we refer the reader for further details. Here we will simply provide a brief summary of the features that will be relevant for our subsequent comparison with data. Throughout this discussion one should bear in mind that PCA is a non-parametric method for constraining the dark energy equation of state. In assessing its performance, one should not compare it to parametric methods. Indeed, no such comparison is possible (even in principle), since the two methods are addressing different questions. Instead one should compare it with another non-parametric reconstruction, and for our purposes with varying couplings the type Ia supernovae provide a relevant comparison.

One can divide the relevant redshift range into $N$ bins such that in bin $i$ the equation of state parameter takes the value $w_i$, 
\begin{equation} 
w(z) = \sum_{i = 1}^N w_i \theta_i(z) \,. 
\end{equation} 
Another way of saying this is that $w(z)$ is expanded in the basis $\theta_i$, with $\theta_1 =(1,0,0,...)$, $\theta_2 = (0,1,0,...)$, etc.

The precision on the measurement of $w_i$ can be inferred from the Fisher matrix of the parameters $w_i$, specifically from $\sqrt{(F^{-1})_{ii}}$, and increases for larger redshift. One can however find a basis in which all the parameters are uncorrelated. This can be done by diagonalizing the Fisher matrix such that $F = W^T \Lambda W$ where $\Lambda$ is diagonal and the rows of $W$ are the eigenvectors $e_i(z)$ or the principal components. These define the new basis in which the new coefficients $\alpha_i$ are uncorrelated and now we can write 
\begin{equation} 
\label{recw} w(z) = \sum_{i = 1}^N \alpha_i e_i(z) \,. 
\end{equation} 
The diagonal elements of $\Lambda$ are the eigenvalues $\lambda_i$ (ordered from largest to smallest) and define the variance of the new parameters, $\sigma^2(\alpha_i) = 1/\lambda_i$.

We will consider the standard class of models for which the variation of the fine-structure constant $\alpha$ is linearly proportional to the displacement of a scalar field, and further assume that this field is a quintessence type field, i.e. responsible for the current acceleration of the Universe \cite{Dvali:2001dd,Chiba:2001er,Anchordoqui:2003ij,Copeland:2003cv,Marra:2005yt,Dent:2008vd}. We take the coupling between the scalar field and electromagnetism to be 
\be 
{\cal L}_{\phi F} = - \frac{1}{4} B_F(\phi) F_{\mu\nu}F^{\mu\nu} ,
\ee 
where the gauge kinetic function $B_F(\phi)$ is linear, 
\be 
B_F(\phi) = 1- \zeta \kappa (\phi-\phi_0) ,\label{coupling}
\ee 
$\kappa^2=8\pi G$ and $\zeta$ is the coupling constant, which in what follows will be marginalized over. This can be seen as the first term of a Taylor expansion, and should be a good approximation if the field is slowly varying at low redshift. Then, the evolution of $\alpha$ is given by 
\begin{equation} 
\frac{\Delta \alpha}{\alpha} \equiv 
\frac{\alpha-\alpha_0}{\alpha_0} = \zeta \kappa (\phi-\phi_0) \,. 
\end{equation} 
For a flat Friedmann-Robertson-Walker Universe with a canonical scalar field, $\dot{\phi}^2 = (1+w(z))\rho_\phi$, hence, for a given dependence of the equation of state parameter $w(z)$ with redshift, the scalar field evolves as 
\begin{equation} 
\phi(z)-\phi_0 = \frac{\sqrt{3}}{\kappa} 
\int_0^z \sqrt{1+w(z)} \left(1+ \frac{\rho_m}{\rho_\phi}\right)^{-1/2} 
\frac{dz}{1+z} . 
\end{equation} 
where we have chosen the positive root of the solution. Note that this allows us to write the evolution of $\alpha$ as
\begin{equation} 
\frac{\Delta\alpha}{\alpha}(z)=\kappa
\int_0^z \sqrt{3[1+w(z)]\Omega_\phi(z)}\frac{dz}{1+z}\,, 
\end{equation} 
where $\Omega_\phi=\rho_\phi/(\rho_m+\rho_\phi)$ is the fraction of the universe's energy in the scalar field.

From the above one can calculate the Fisher matrix using standard techniques, as discussed in \cite{Amendola}. As in that work, we will consider three fiducial forms for the equation of state parameter:
\begin{equation}
w_{c}(z)=-0.9,
\end{equation}
\begin{equation}
w_{s}(z)=-0.5+0.5 \tanh \left(z-1.5 \right),
\end{equation}
\begin{equation}
w_{b}(z)=-0.9+1.3 \exp{ \left[-\frac{(z-1.5)^2}{0.1} \right]}\,.
\end{equation}
At a phenomenological level, these describe the three qualitatively different interesting scenarios: an equation of state that remains close to a cosmological constant throughout the probed redshift range, one that evolves towards a matter-like behavior by the highest redshifts probed, and one that has non-trivial features over a limited redshift range, perhaps associated to a low-redshift phase transition (see \cite{Mortonson} for further discussion). Thus in what follows we will refer to these three cases as the \textit{constant}, \textit{step} and \textit{bump} fiducial models.

We will assume a flat universe, and further simplify the analysis by fixing $\Omega_m=0.3$. This is a standard procedure, that was followed in the original paper of Huterer and Starkman  \cite{Huterer:2002hy} and also in a number of subsequent works. This specific choice of $\Omega_m$ has a negligible effect on the main result of the analysis, which is the uncertainty in the best determined modes. For each fiducial model we choose the coupling such that it leads to a few parts-per-million variation of $\alpha$ at redshift $z\sim4$, consistently with \cite{Dipole}. In \cite{Amendola} the analysis was focused on forecasts for ESPRESSO \footnote{See http://www.eso.org/sci/facilities/develop/instruments/espresso.html} and CODEX---now dubbed ELT-HIRES \footnote{See http://www.hires-eelt.org/}. Here we will start by discussing a more general analysis, leaving specific choices to a later section.

In order to systematically study possible observational strategies, it is of interest to find an analytic expression for the behavior of the uncertainties of the best determined PCA modes described above. For this one needs to explore the range of parameters such as the number of $\alpha$ measurements ($N_\alpha$) and the uncertainty in each measurement ($\sigma_\alpha$). For simplicity we will assume that this uncertainty is the same for each of the measurements in a given sample, and also that the measurements are uniformly distributed in the redshift range under consideration.

By exploring numbers of measurements $N_\alpha$ between 20 and 200, uniformly distributed in redshift up to $z=4$, and individual measurement uncertainties between $10^{-5}$ and $10^{-8}$ we find the following fitting formula for the uncertainty $\sigma_n$ for the $n$-th best determined PCA mode
\begin{equation}\label{theoryfit}
\sigma_n=A\frac{\sigma_\alpha}{N_\alpha^{0.5}}[1+B(n-1)]\,.
\end{equation}
The coefficients A and B will depend on the choice of fiducial model, and also on the number of PCA bins assumed for the redshift range under consideration. Table \ref{table1} lists these coefficients for choices of 20 and 30 bins. Notice that it is useful to provide the uncertainly $\sigma_\alpha$ in the fitting formula in parts per million, since in that case the coefficients $A$ and $B$ are of order unity.

\begin{table}
\begin{tabular}{|c|c|c|c|c|}
\hline
Model & A ($N_{\rm b}=20$) & B ($N_{\rm b}=20$) & A ($N_{\rm b}=30$) & B ($N_{\rm b}=30$)  \\
\hline
Constant & 1.14 & 0.52 & 1.39 & 0.63 \\
Step & 2.10 & 0.96 & 2.53 & 1.16 \\
Bump & 1.65 & 0.75 & 2.00 & 0.91 \\
\hline
\end{tabular}
\caption{\label{table1} The coefficients A and B in the fitting formula \protect\ref{theoryfit}, assuming $N_{\rm b}=20$ (left side of the table) and $N_{\rm b}=30$ (right side) PCA bins in the redshift range $0<z<4$ and uncertainties $\sigma_\alpha$ expressed in parts per million.}
\end{table}

A comparison between the numerically determined values and our fitting formula indicates that for $N_\alpha>50$ the present expression is reasonably accurate for all values up to and including $n=6$, while for a smaller number of measurements the number of accurately determined modes is less than 6 (for example for$N_\alpha=20$ only the first two modes obey the above relation, with the uncertainty in next two being slightly higher than suggested by the formula---and that of the next two significantly so. Speifically, Table \ref{tablex} shows the average and maximal relative error obtained by sampling the above parameter space of $(\sigma_\alpha,N_\alpha,n)$, for a fixed number of redshift bins $N_{\rm b}=20$. The maximal errors always occur for high $n$ and low $N_\alpha$, while in the opposite corner of parameter space they are below $10\%$. By smapling uniformly in $N_\alpha$ and in the logarithm of $\sigma_\alpha$ one obtains averaage uncertainties around $30\%$, which are adequate considering the simplifying assumptions in our modelling.

\begin{table}
\begin{tabular}{|c|c|c|}
\hline
Model & Average Error & Max. Error  \\
\hline
Constant & $29\%$ & $38\%$ \\
Step & $37\%$ & $48\%$ \\
Bump & $26\%$ & $37\%$ \\
Average & $51\%$ & $67\%$ \\
\hline
\end{tabular}
\caption{\label{tablex} The average and maximal errors of our fitting formula \protect\ref{theoryfit}, compared to the correct PCA result. We have assumed $N_{\rm b}=20$. The first three lines show the results for aech of the three fiducial models, while the fourth line shows the result of trying to describe all three models with a single 'average' fitting formula, where the values of coefficients A and B are the averages of those for the individual models.}
\end{table}

Overall, the fitting formulae show some dependence on the specific model being considered. One may ask if by taking say the arithmetic mean of the values of the coefficients A and B for the three models one will obtain a generic fitting formula that will be reasonable for all three. The last line of Table \ref{tablex} shows that this is not the case, as the uncertainties worsen considerably: the average values of A and B ara quite close to those of the bump fiducial model, but these coefficients do not perform as well for the other models. This model-dependence should therefore be taken into consideration if we want to establish a simple optimization pipeline, since the correct redshift evolution of the dark energy equation of state is not known \textit{a priori} (certainly not at the high redshifts that can be probed thorough this method). There is also dependence on the number of bins, which is to be expected: as we increase the number of bins the uncertainties in each bin will increase. Despite these caveats, the fitting formulas, once further calibrated using actual data (as will be done in the next section) will allow us to quantify the ability of a particular spectrograph to distinguish between different models.


\section{\label{uves}Calibrating the fitting formula with VLT data}

The next step is then to connect these theoretical tools to observational specifications. A time normalization can in principle be derived from the present VLT performances, with the caveat that the present errors on $\alpha$ are dominated by systematics and not by photons. Nevertheless, we can assume a simple (idealized) observational formula,
\begin{equation}\label{naivefit}
\sigma_{sample}^{2}=\frac{C}{T}\,,
\end{equation}
where $C$ is a constant, $T$ is the time of observation necessary to acquire a sample of spectra from which one will obtain $N$ measurements of $\alpha$ at the relevant redshifts, and $\sigma_{sample}$ is the relative uncertainty in these measurements (ie, the uncertainty in $\Delta\alpha/\alpha$) for the whole sample. This is expected to hold for a uniform sample, meaning a sample with $N_\alpha$ identical objects, each of which produces a measurement with the same uncertainty $\sigma_\alpha$ in a given observation time. Naturally any real-data sample will not be uniform, so there will be some corrections to this behavior. The uncertainty of the sample will be given by
\begin{equation}
\sigma_{sample}^{2} =\frac{1}{\sum^{N}_{i=1} \sigma_{i}^{-2}}\,,
\end{equation}
and for the above simulated case with $N$ measurements all with the same $\alpha$ uncertainty we simply have
\begin{equation}
\sigma_{sample}^{2} =\frac{\sigma_\alpha^2}{N}\,.
\end{equation}

Clearly there are also other relevant observational factors that a simple formula like this does not take into account, in particular the structure of the absorber (the number and strength of the components, and how narrow they are) and the position of the lines in the CCD, which is connected to the redshift of the absorption system. The latter is also related to the wavelength range covered by each spectrograph. A further issue (which is easier to deal with) is the fact that a given line of sight often has several absorption systems, and thus yields several different measurements. Despite these caveats, this formula is adequate for our present purposes, as will be further discussed below.

We have used the UVES data from Julian King's PhD thesis \cite{kingthesis}, complemented by observation time data provided by Michael Murphy, to build a sample to calibrate the observational formula. In addition to these properties of the dataset, we also calculated the signal to noise per pixel with the following equation, parametrized by Michael Murphy using specifications of UVES spectrograph:
\begin{equation} \label{eq:SNR}
SNR = K \left[\frac{T}{T_{0}} 10^{-0.4(M-M_{0})}\right] ^{1/2}
\end{equation}
where $T$ is the exposition time, $M$ is the magnitude of the source and for $K=20$, $T_{0}=3600s$ and $M_{0}=17.8$. However, we note that this SNR is calculated for illustration purposes only, and is not used in our fitting analysis.

\begin{figure}
\includegraphics[width=\hsize]{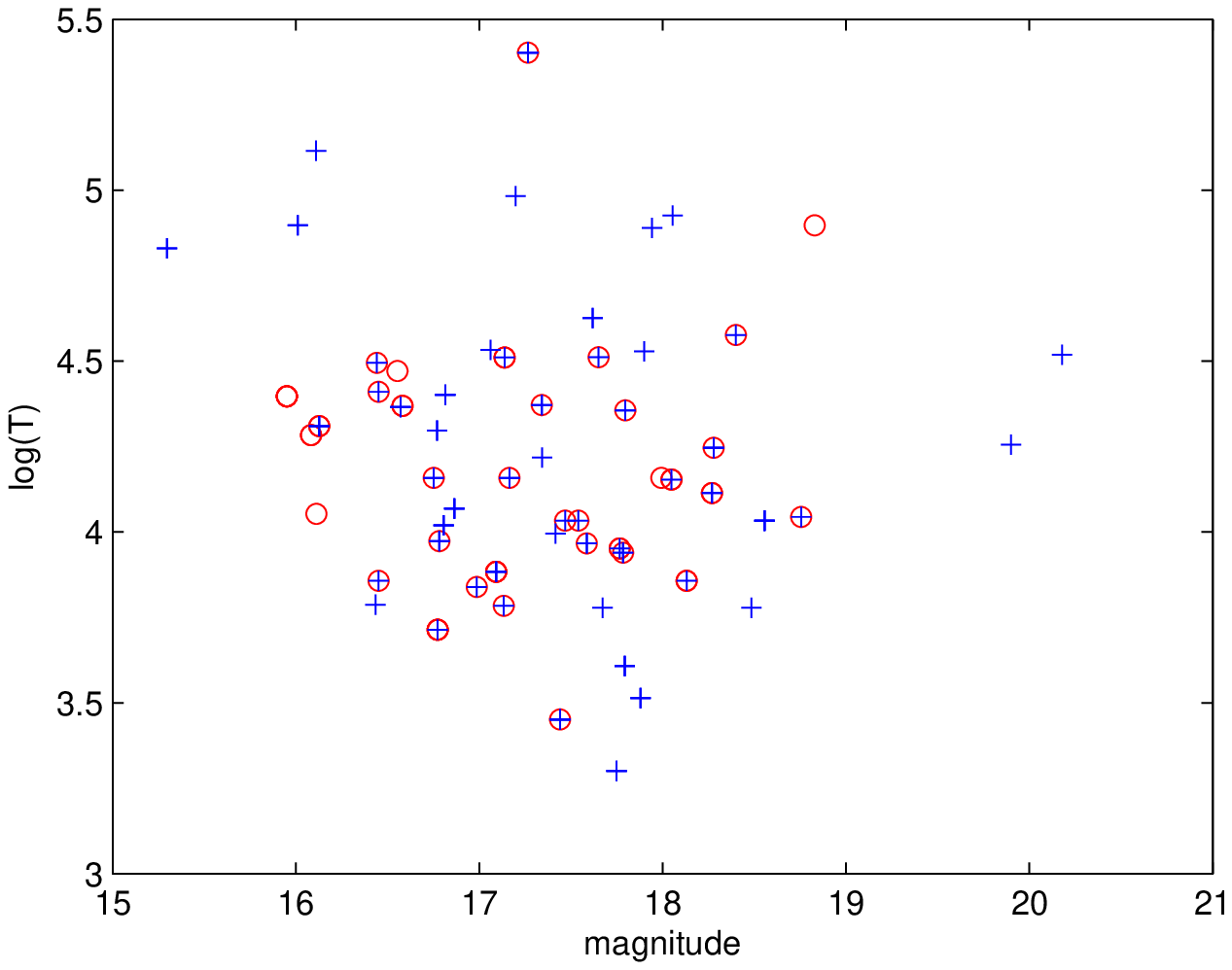}
\includegraphics[width=\hsize]{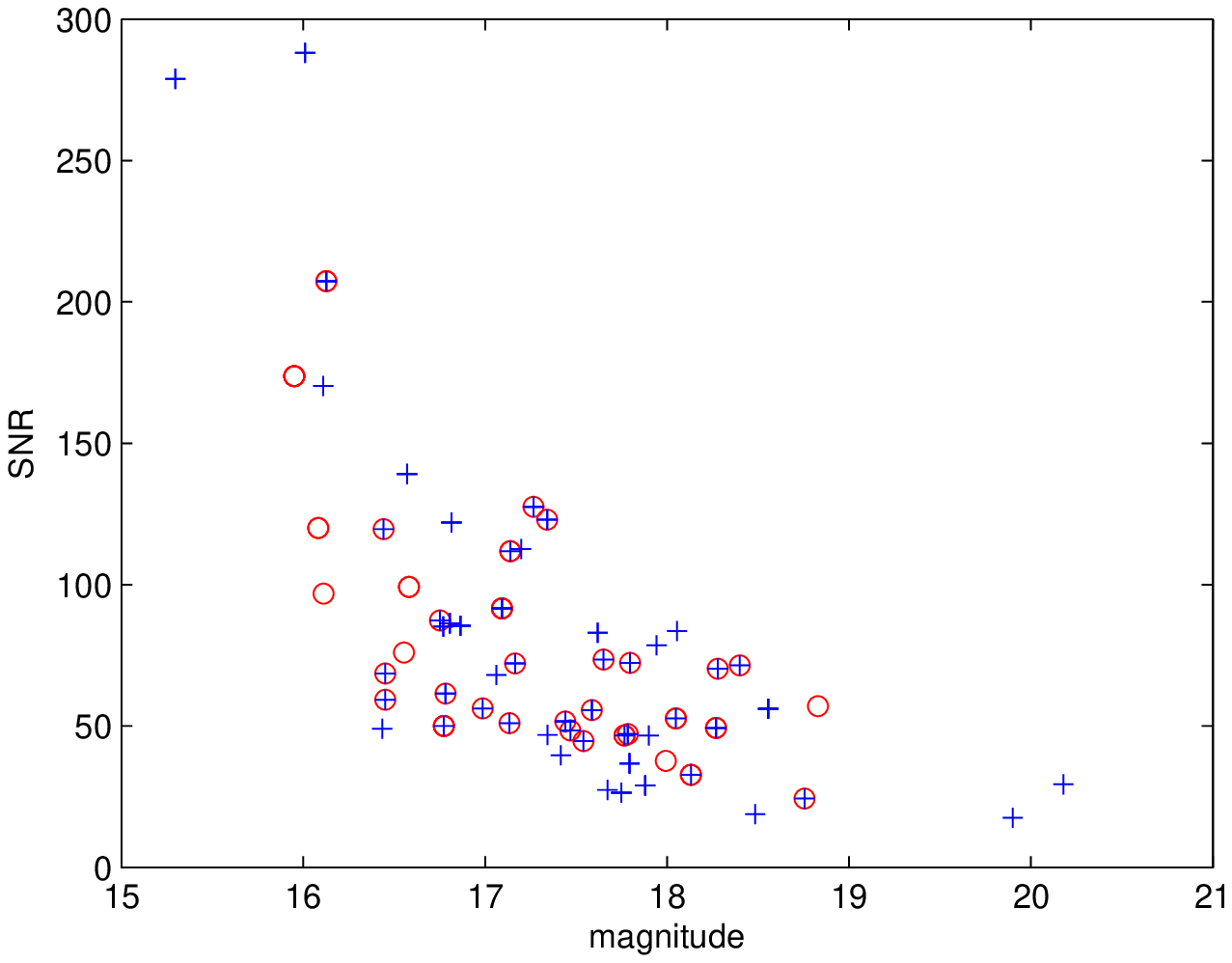}
\caption{\label{corrmag}Observation time and SNR for the VLT absorbers of \protect\cite{kingthesis}, as a function of the quasar magnitude. Circles denote absorbers yielding measurements with better than 10 ppm statistical uncertainty, crosses denote the rest of the absorbers.}
\end{figure}

\begin{figure}
\includegraphics[width=\hsize]{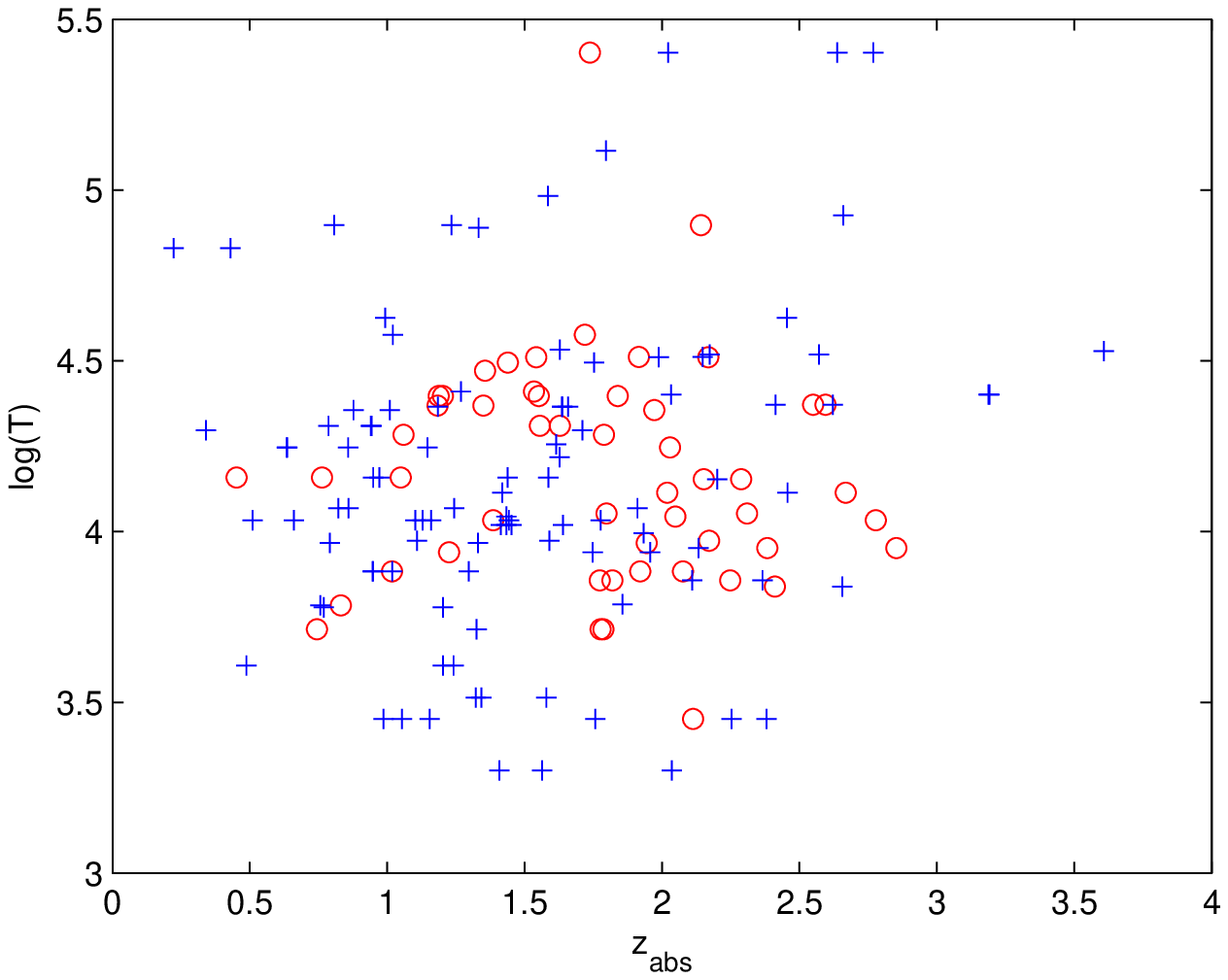}
\includegraphics[width=\hsize]{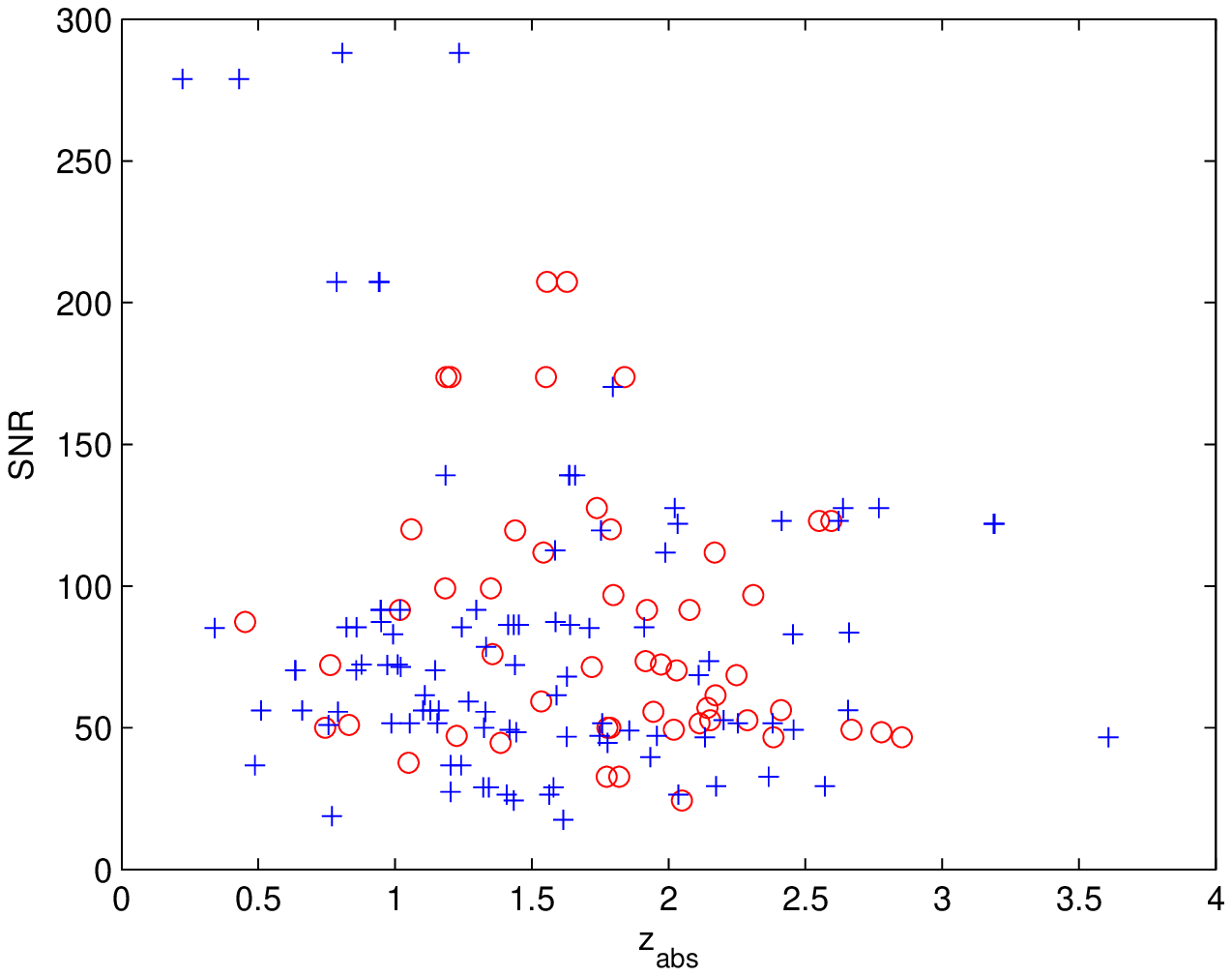}
\caption{\label{corrzabs}Observation time and SNR for the VLT absorbers of \protect\cite{kingthesis}, as a function of the redshift of the absorption system. Circles denote absorbers yielding measurements with better than 10 ppm statistical uncertainty, crosses denote the rest of the absorbers.}
\end{figure}

\begin{figure}
\includegraphics[width=\hsize]{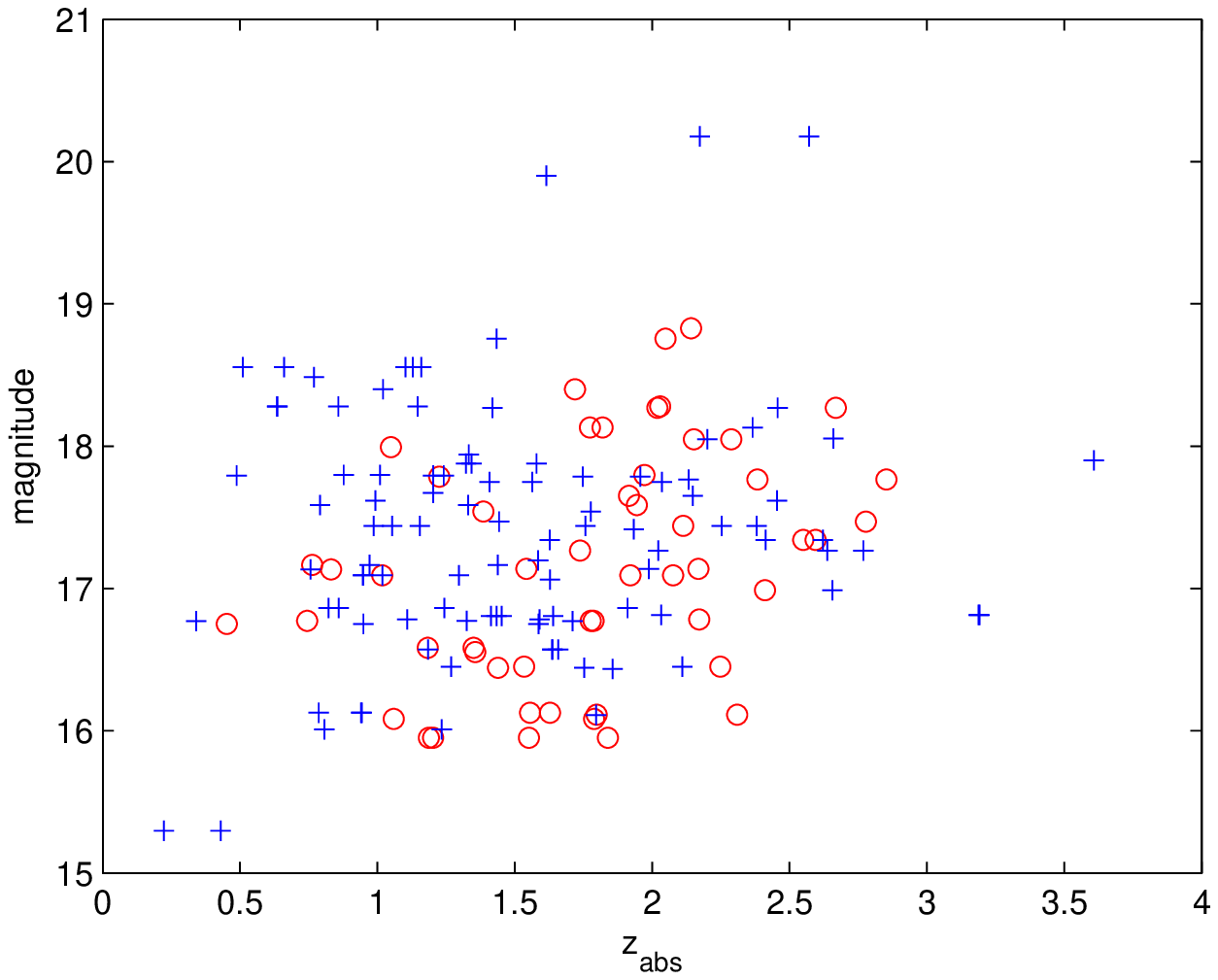}
\caption{\label{corrmix}Uncertainty in the $\alpha$ measurements for the VLT absorbers of \protect\cite{kingthesis}, as a function of the magnitude of the quasar and the redshift of the absorbers. Circles denote absorbers yielding measurements with better than 10 ppm statistical uncertainty, crosses denote the rest of the absorbers.}
\end{figure}

Figs. \ref{corrmag}, \ref{corrzabs} and \ref{corrmix} display some relevant properties of this set of absorption systems, including the magnitude of the quasar, the redshift of the absorber, the observation time and the SNR of the spectrum. In all cases the circles denote the absorbers that lead to measurements with better than 10 parts per million statistical uncertainty, whereas crosses depict the rest of the absorbers. Note that several lines of sight contain multiple absorption systems, which is why several circles and crosses overlap in the magnitude-time panel of Fig. \ref{corrmag}.

It is clear that this sample is far form ideal, as it does not display the types of correlations that one would expect from such a sample: better SNR or observation time do not necessarily lead to a better measurement of $\alpha$. Undoubtedly this is a consequence of having a dataset put together from archival data. We do find the obvious correlation between SNR and the magnitude of the quasar (bottom panel of Fig. \ref{corrmag}). The more interesting result of this analysis is shown in Fig. \ref{corrmix}, which shows that higher redshift absorbers lead to proportionally better measurements. Moreover, in low-redshift absorbers brighter systems tend to give better measurements, while for higher redshift ones fainter systems can still yield good measurements. The reason for these differences stems form the different transitions within the range of the spectrograph at the various redshifts---see \cite{kingthesis} for further discussion.


\begin{figure}
\includegraphics[width=\hsize]{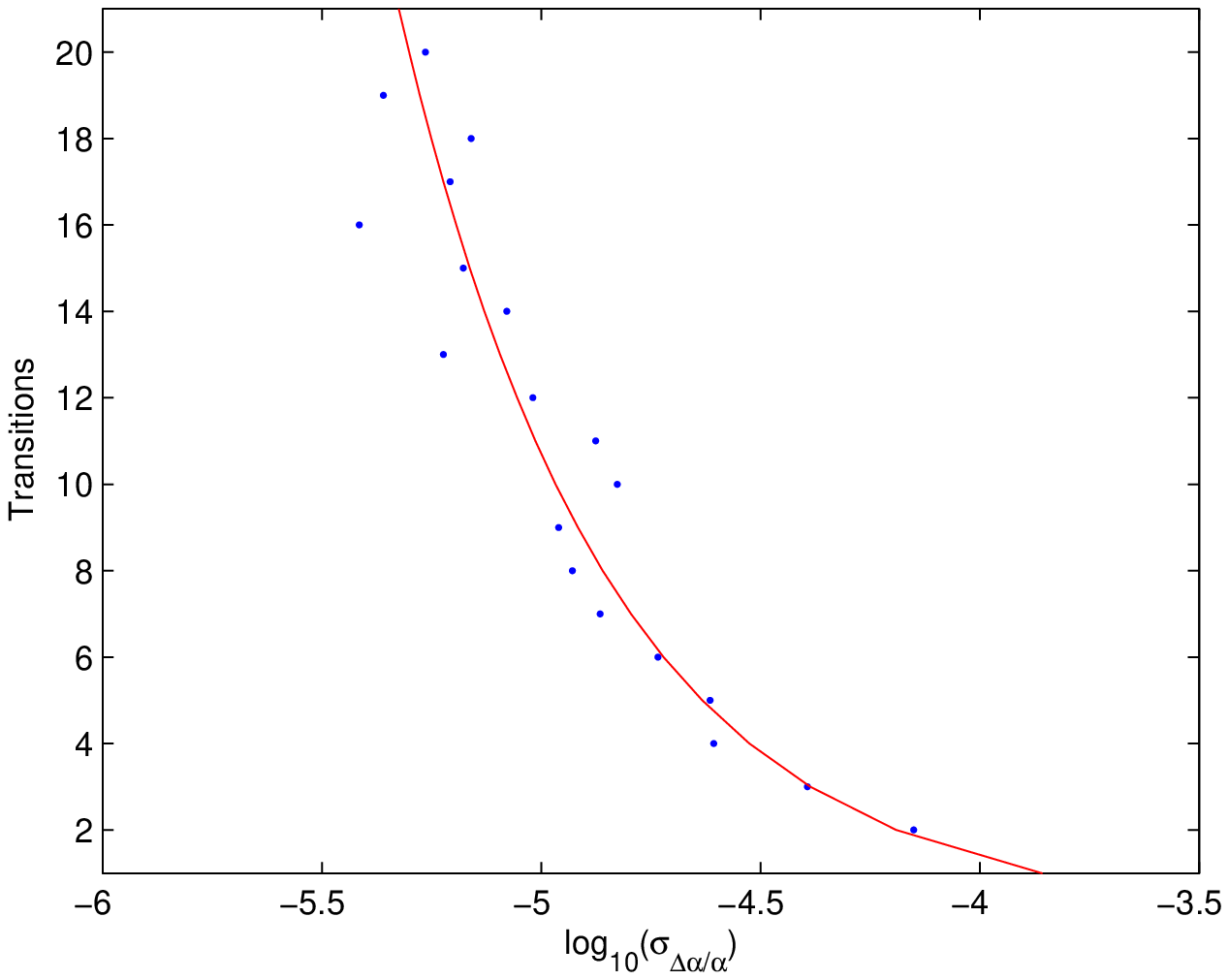}
\caption{\label{numlines}Correlation between statistical uncertainty of each of the $\alpha$ measurements and the number of transitions used to obtain them. Each point in the plot was obtained as an average of the various points in the dataset with each number of transitions used. The red line is the best polynomial fit, discussed in the text.}
\end{figure}

We do find a strong correlation between the number of transitions used to make one measurement ($N_{\lambda}$) and the statistical uncertainty of the measurement, as can be observed in Fig. \ref{numlines} where, for each $N_\lambda$, we plot the average uncertainty in the $\alpha$ measurements, $\sigma_{\Delta\alpha/\alpha}$, achieved as a function of that number of transitions. (Note that these transitions need not be the same is the various cases being averaged over.) We find that a simple parametrization shows the following approximate relation
\begin{equation}\label{fig:numlines}
\sigma_{\Delta \alpha / \alpha} = 139 N_{\lambda}^{-1.11}\, {\rm ppm},
\end{equation}
where again we expressed the uncertainty in part per million.
This best-fit parametrization is also plotted in Fig. \ref{numlines}.

In passing, we note that there is also a correlation between the sensitivity of the measurements and the (absolute) value of the $q$-coefficients of the transitions being used. This is unsurprising: transitions that shift the most for a given shift in $\alpha$ tend to yield better measurements. However, we shall not quantify this correlation, since it does not directly impact the phenomenological modelling of the present work.

One consequence of these non-ideal properties of the sample is that the simple relation given by Eq. (\ref{naivefit}) will not strictly hold. Nevertheless, there is a simple way to correct it, which consists of allowing the former constant $C$ to itself depend on the number of sources. This is easy to understand: in a small sample one typically will have the best available sources; by increasing our sample we will be adding sources which are not as good as the previous ones, and therefore the overall uncertainty in the $\alpha$ measurement will improve more slowly than in the ideal case---or alternatively one will need additional telescope time to do so.

\begin{figure}
\includegraphics[width=\hsize]{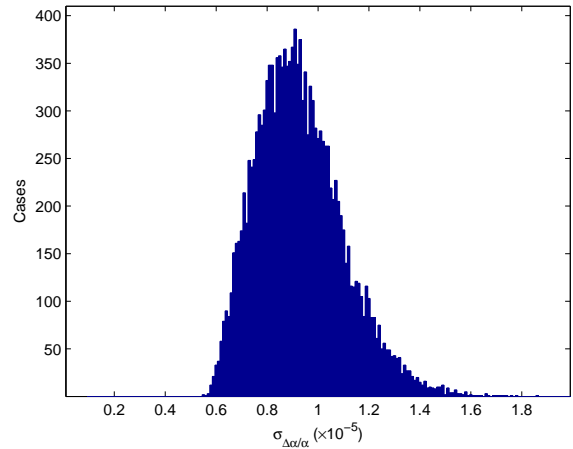}
\includegraphics[width=\hsize]{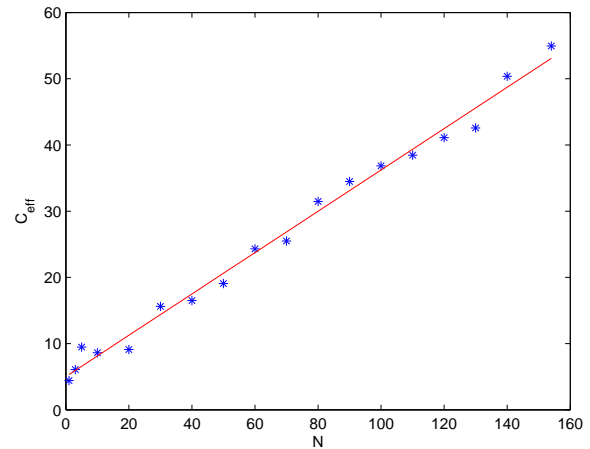}
\caption{\label{NCeff}Top: Distribution of uncertainties in $\alpha$ for 20-source VLT subsamples, for a total of 15000 realizations. Bottom: Values of the effective parameter $C$ as a function of the number of systems considered, for the parametrisation of the observational formula applied to the current UVES data. The red line is the best linear fit, discussed in the text.}
\end{figure}

Using standard Monte Carlo techniques we have generated several tens of thousands of sub-samples of the VLT sample, with various numbers of sources, for which we determined the overall uncertainty in the $\alpha$ measurement and the amount of telescope time needed to achieve it. From these distributions (an example of which, for the case $N=20$, is shown in the top panel of Fig. \ref{NCeff}) one can determine the corresponding mean values, and these then allow us to infer the behavior for the empirical function $C(N)$. The results of this analysis are shown in Fig. \ref{NCeff}. We find that a good fit is provided by the linear relation
\begin{equation}\label{cuves}
C(N_\alpha)=0.31\, N_\alpha + 5.02\,.
\end{equation}

Here the constant has been normalised such that $\sigma_{sample}$ is given in parts per million and $T$ is in nights. As a simple check, for the UVES Large Program for Testing Fundamental Physics \cite{LP1,LP2}, with about 40 nights and 16 sources, we infer from the fitting formula a value of $0.5$ parts per million, consistent with the expectations of the collaboration \cite{LP3}.

Finally, if we add a 'systematics' term $\sigma^2_{sys}$ to Eq. (\ref{naivefit})  and repeat the above procedure, our simple analysis indicates that values
\begin{equation}
\sigma_{sys}\sim 4-6\, ppm\,
\end{equation}
provide a reasonable fit. It is interesting to note that this is not too distant from the value obtained in \cite{Dipole,kingthesis},
\begin{equation}
\sigma_{Webb}=9\, ppm\,;
\end{equation}
naturally, their value was obtained with a much more sofisticated analysis. Nevertheless, this suggests that our simple toy modelling does capture the salient broad features of the datasets.


\section{\label{future}Future observational facilities}

We can now put together the results of the two previous sections to obtain a UVES-calibrated PCA formula
\begin{equation}
\sigma_n=A[1+B(n-1)]\frac{\sigma_\alpha}{\sqrt{N_\alpha}}=A[1+B(n-1)]\left[\frac{C(N_\alpha)}{T}\right]^{1/2}\,,
\end{equation}
where the UVES $C(N)$ formula is given by Eq. (\ref{cuves}). The most striking feature of this result is the explicit (and strong) dependence on the number of sources. Future improvements will come from a better sample selection and optimized acquisition/calibration methods and both of these are expected to significantly reduce this dependence, even elliminating it for moderately sized samples of absorbers. In the case of the ELT-HIRES, a further improvement will come from the larger collecting power.

\begin{figure}
\includegraphics[width=\hsize]{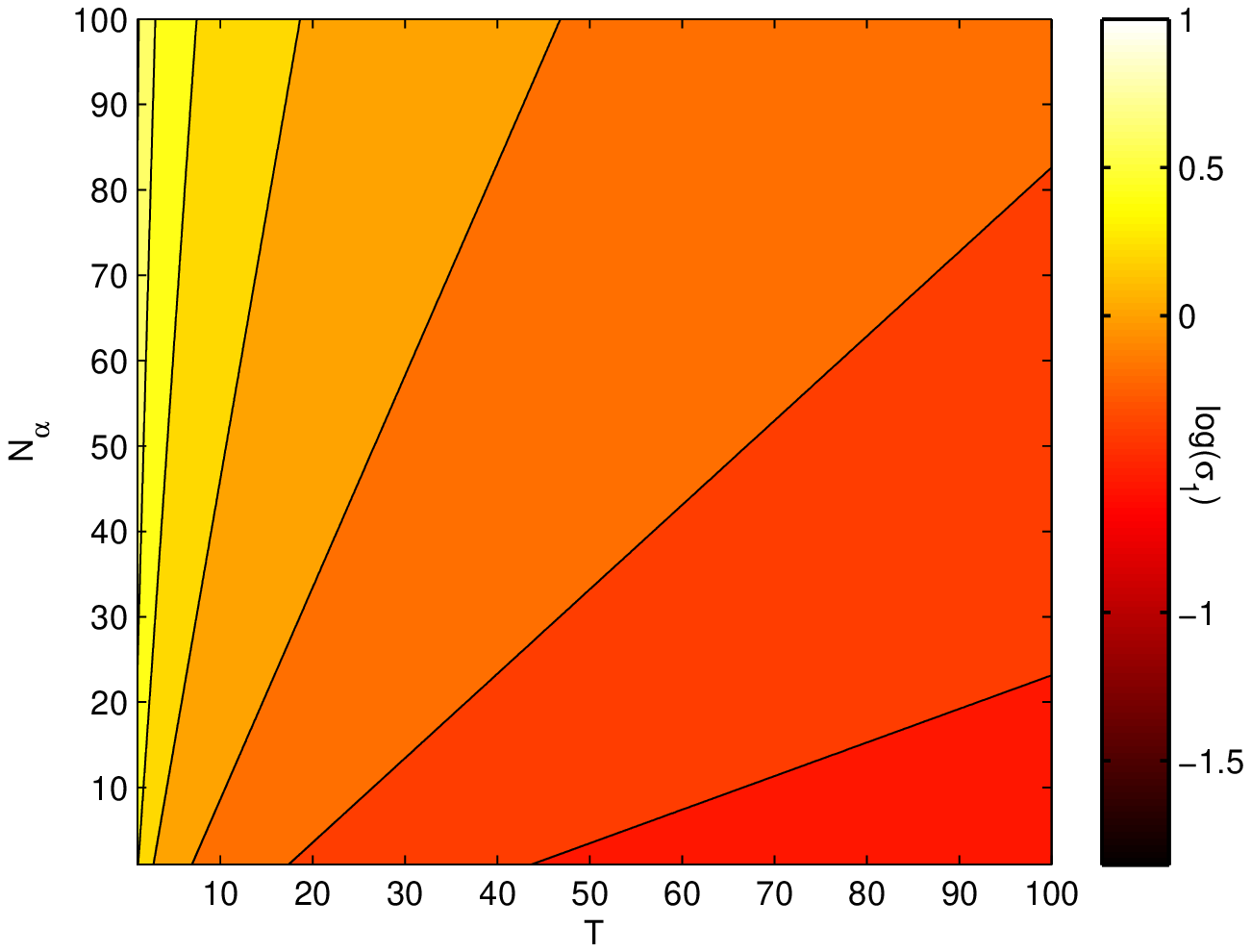}
\includegraphics[width=\hsize]{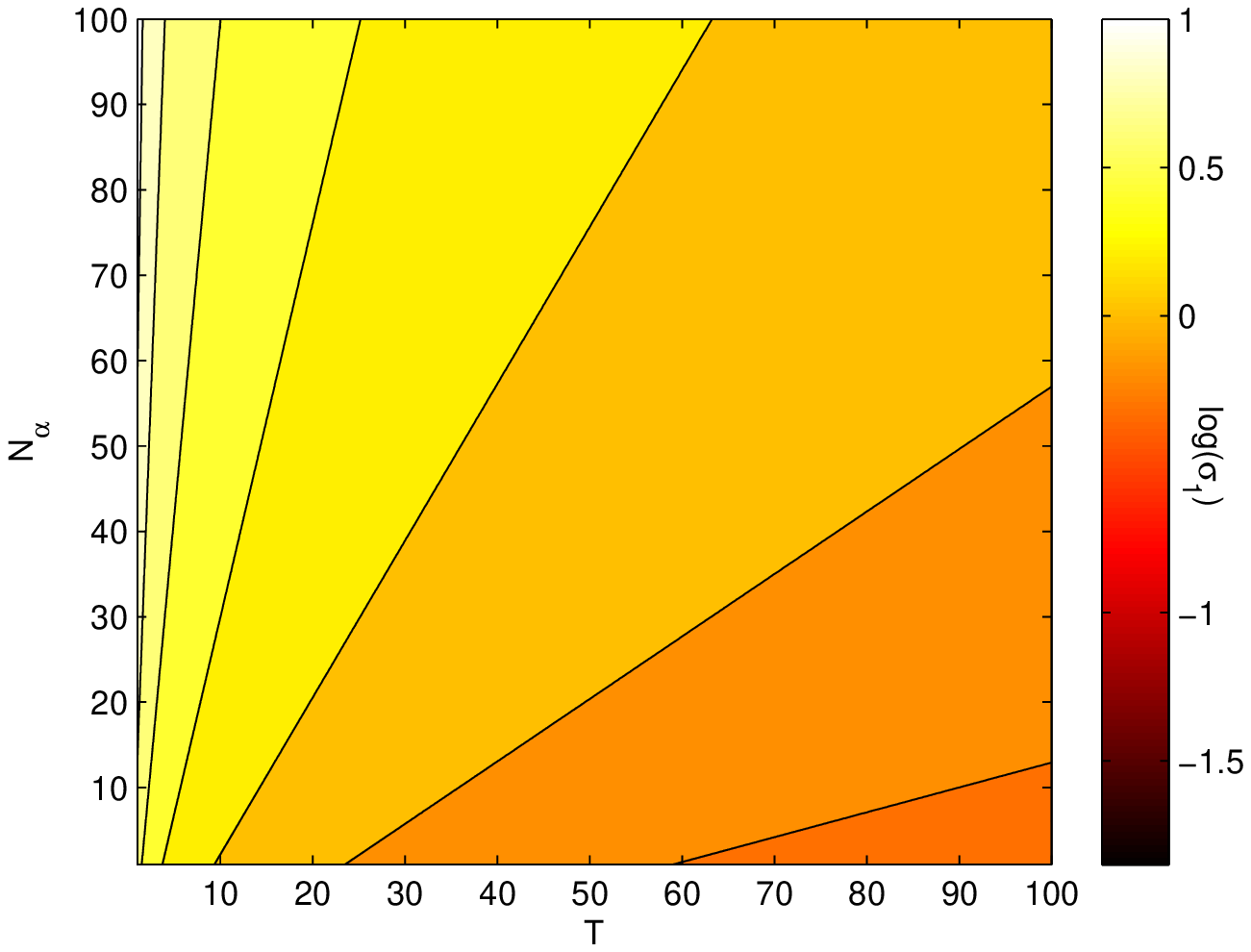}
\includegraphics[width=\hsize]{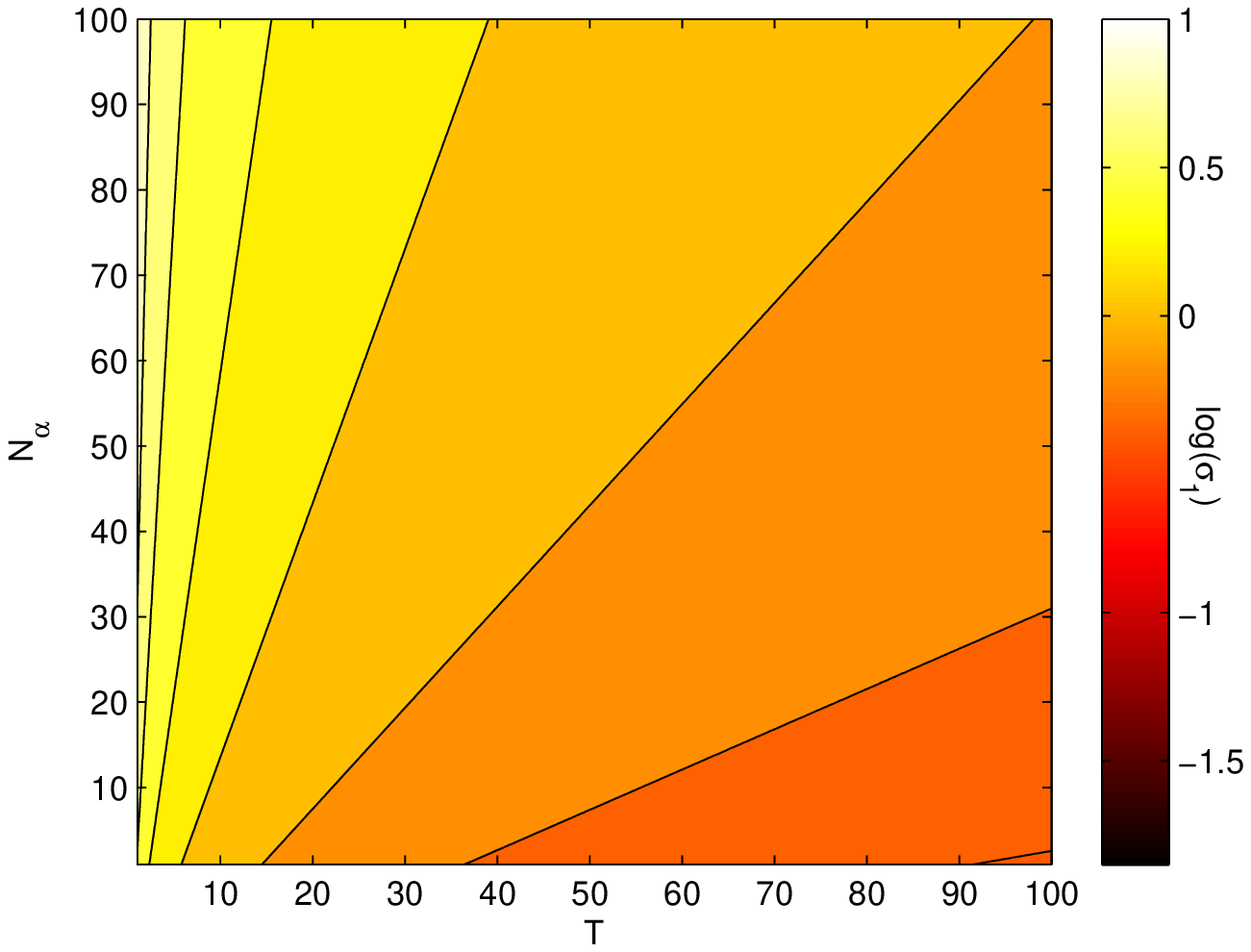}
\caption{\label{esprebase}The uncertainty in the best-determined PCA model in baseline scenario described in the main text, as a function of the number of nights of observation and absorbers measured, respectively for the constant, step and bump fiducial models (top to bottom). In each case the colormap indicates the logarithm of the uncertainty.}
\end{figure}

With simple but reasonable extrapolations we can forecast the expected changes to the UVES formula, and from this carry out an assessment of the impact of these measurements for constraining dark energy. We shall consider three scenarios
\begin{itemize}
\item A baseline scenario, where there are essentially no improvements over UVES, that is
\begin{equation}\label{cesp1}
C(N_\alpha)_{BASE}=0.31\, N_\alpha + 5.02\,;
\end{equation}
this reflects the current situation, and therefore provides a benchmark against which future improvements can be discussed. Note that although this phenomenological fitting formula was obtained for UVES at the VLT, we expect it to also apply---at least qualitatively---to analogous contemporary spectrographs in other 8-meter class telescopes, such as HIRES-Keck or HDS-Subaru,
\item An ESPRESSO scenario, where
\begin{equation}\label{cesp2}
C(N_\alpha)_{ESPRESSO}=\frac{5.02}{9}\,;
\end{equation}
given realistic estimates of the available time (note that 27 GTO nights are currently foreseen) the observable samples are small enough to make a factor of 3 gain (on average) in sensitivity due to improved signal-to-noise and resolution, while elliminating the explicit dependence of $C$ on the number of sources. These improvements arise from the fact that it will be, by design \cite{ESP0}, free of the systematics that are known to affect UVES, and in particular to the much more precise wavelength calibration, which will be done with a Laser Frequency Comb. Note that ESPRESSO does have a wavelength coverage that is substantially reduced compared to that of UVES, and this will certainly offset some of the above improvements.
\item An ELT-HIRES scenario, where
\begin{equation}\label{chires}
C(N_\alpha)_{HIRES}=\frac{5.02}{300}\,;
\end{equation}
here we similarly expect a constant $C$ parameter (even allowing for the larger number of absorbers measured), and further gains in sensitivity have been factored in, including the five-fold increase in the telescope collecting area. Another key advantage of ELT-HIRES is its wide wavelength coverage, not only in the ultraviolet and optical but also in the infrared.
\end{itemize}

\begin{figure}
\includegraphics[width=\hsize]{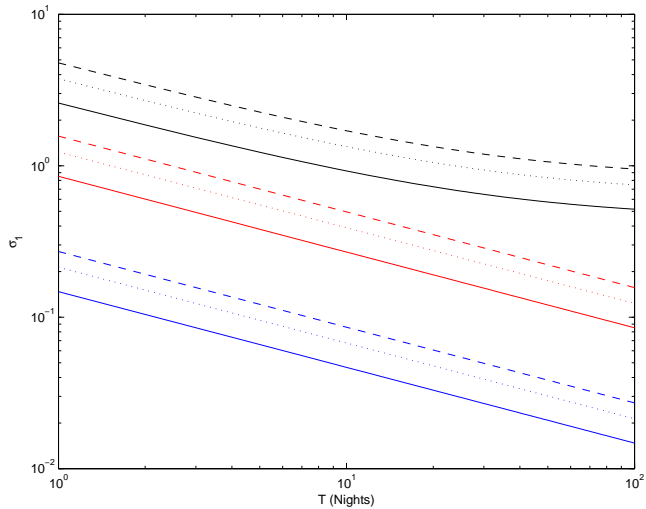}
\caption{\label{threecase}The uncertainty in the best-determined PCA mode in the three scenarios discussed in the main text, for each of the fiducial models considered. The top (black), middle (red) and bottom (blue) sets of three lines correspond to the baseline, ESPRESSO and ELT-HIRES cases respectively. In each set the solid, dashed and dotted lines respectively correspond to the constant, step and bump fiducial models.}
\end{figure}

Figs. \ref{esprebase} and \ref{threecase} depict the uncertainty in the best-determined PCA mode, for the three observational scenarios discussed above and the three fiducial models considered (the constant, step and bump models). In these, and throughout the discussion in this section, we will assume 20 PCA bins ($N_b=20$, cf. Table \ref{table1}). The former figure highlights the dependence on the number of sources in the baseline scenario, while the latter figure highlights the gains to be expected from ESPRESSO and ELT-HIRES. For the baseline scenario in this latter plot we assumed a number of sources equal to half the number of nights, which is a typical number for current observations.

\begin{table}
\begin{tabular}{|c|c|c|c|}
\hline
Model & Baseline & ESPRESSO & ELT-HIRES \\
\hline
Constant & 8.2 & 0.7 & 0.02 \\
Step & 70.0 & 2.5 & 0.07 \\
Bump & 23.6 & 1.5 & 0.05 \\
\hline
\end{tabular}
\caption{\label{table2}Number of nights needed to achieve an uncertainty of unity in the best-determined PCA mode, $\sigma_1=1$, for the various scenarios and fiducial models considered. For the baseline scenario $N_\alpha=T/2$ was assumed.}
\end{table}

\begin{table}
\begin{tabular}{|c|c|c|}
\hline
Model & ESPRESSO & ELT-HIRES \\
\hline
Constant & 649.8 & 19.5 \\
Step & 2231.6 & 66.9 \\
Bump & 1420.1 & 42.6 \\
\hline
\end{tabular}
\caption{\label{table3}Number of nights needed to achieve, with $\alpha$ measurements uniformly spaced in redshift, an uncertainty in the best-determined PCA mode equal to that expected from a SNAP-like dataset of 3000 Type Ia supernovas, for the ESPRESSO and ELT-HIRES scenarios and the various fiducial models considered. Note that this is not possible at all in the baseline scenario}
\end{table}

An alternative way to quantify the expected improvements with ESPRESSO and ELT-HIRES is to estimate the number of observation nights needed to obtain an uncertainty in the best-determined PCA mode of $\sigma_1=1$. This is shown in Table \ref{table2}, where we again assumed $N_\alpha=T/2$ for the baseline scenario, and the gains are obvious. Note that here the model-dependence is enhanced, since the observation time will depend on the square of the coefficent $A$.

For a more ambitious goal, we can instead estimate the number of nights needed to reach the same sensitivity on the first PCA mode as `SNAP-like' dataset of 3000 supernovas. This turns out to be $\sigma_{1,SNAP}\sim0.033$, with the model dependence appearing at the next decimal place \cite{Amendola}. In this case we find that this level of sensitivity is not achievable at all with current facilities, while our estimates for ESPRESSO and ELT-HIRES are listed in Table \ref{table3}. Importantly we see that a few tens of nights are sufficient for ELT-HIRES, further highlighting the key role that the ELT will be able to play on fundamental cosmology.

We note that a uniform redshift cover is important in obtaining these results. Moreover the range of redshifts considered will also play a role, as it will determine how many useful transitions will fall within the range of the spectrograph. A more detailed study of these effects is left for future work.


\section{\label{concl}Conclusions}

We have highlighted how the forthcoming generation of high-resolution ultra-stable spectrographs will play a crucial role in the ongoing search for the new physics that is currently powering the acceleration of the universe, We focused on ongoing and planned astrophysical tests of the stability of nature's fundamental couplings, specifically discussing the improvements that can be expected with ESPRESSO and ELT-HIRES and their impact on fundamental cosmology. However, much of what has been said is also relevant for other forthcoming instruments, such as PEPSI at the LBT or HROS at the TMT.

Our analysis suggests different observational strategies for ESPRESSO and ELT-HIRES. In fact that of ELT-HIRES is easy to outline: given its exquisite sensitivity, it should focus on mapping out the behavior of $\alpha$ on a wide range of redshifts, leading to competitive constraints on dark energy and fundamental physics paradigms. Nevertheless, the choice of redshift ranges to probe may be influenced by the earlier ESPRESSO results. For ESPRESSO, the gains in sensitivity are partially offset by its relatively limited wavelength range, which will limit the range of redshifts that can be mapped at high sensitivity. Although this issue warrants further study, our results suggest that one should concentrate on testing the stability of fundamental couplings using a relatively small set of carefully chosen absorbers. 

Our findings are directly relevant for the target selection process for both spectrographs, and even for the ELT-HIRES Phase A studies, which has clear potential for being a leding instrument in the field of fundamental cosmology. Although we have not specifically addressed the issue of redshift coverage (which we leave for future work), it is clear that a large redshift lever arm for the measurements is important, leading to the requirement of a broad wavelength range for the spectrograph (which also maximizes the number of transitions available for the measurements).

Finally, let us point out that if varying fundamental couplings are confirmed by
ESPRESSO and ELT-HIRES, these spectrographs can themselves carry out consistency tests by looking for additional observational effects that must exist if constants vary. One example, to which both spectrographs can contribute, are tests of the redshift dependence of the cosmic microwave background temperature \cite{Tasos1,Tasos2}. A second example is provided by the redshift drift \citep{sandage,pauline1}, which is probably outside the reach of ESPRESSO but will be a key driver for ELT-HIRES (and may also be measured, at lower redshifts, by other facilities such as the SKA).

\begin{acknowledgments} 
We are grateful to Stefano Cristiani, Paolo Molaro, Michael Murphy and John Webb for many interesting discussions and suggestions. Special thanks to Michael Murphy for providing us with the observation times for the VLT sources.

This work was done in the context of grant PTDC/FIS/111725/2009 from FCT (Portugal). C.J.M. is also supported by an FCT Research Professorship, contract reference IF/00064/2012, funded by FCT/MCTES (Portugal) and POPH/FSE (EC). N.J.N. is also supported by the grants EXPL/FIS-AST/1608/2013 and OE/FIS/UI2751/2014.
\end{acknowledgments}

\bibliography{strategies}

\end{document}